\documentclass[twocolumn,showpacs,preprintnumbers,amsmath,amssymb]{revtex4}
\usepackage{graphicx}
\usepackage{dcolumn}
\usepackage{bm}

\begin{document}

\title{Electronic Structure and Magnetic Properties of $\beta$-Ti$_{6}$Sn$_{5}$}

\author{T. Jeong}

\affiliation{
Department of Physics, University of California, Davis, California 95616
}


\begin{abstract}
The electronic structure of $\beta$-Ti$_{6}$Sn$_{5}$ has been 
studied based on the density functional theory within the 
local-density approximation.
The calculation indicates that $\beta$-Ti$_{6}$Sn$_{5}$ is very 
close to ferromagnetic instability and shows ferromagnetic ordering 
after rare earth element doping. Large enhancement of the static 
susceptibility over its non-interacting value is found due to 
a peak in the density of states at the Fermi level.

\end{abstract}


\maketitle

\section{Introduction}
Electronic systems close to magnetic instability has drawn
a lot of both experimental and theoretical attention because of 
their interesting and often unexpected properties arising from the interaction
of electrons with the magnetic fluctuations. 
Recently close-to-ferromagnetic behavior was observed in $\beta$-Ti$_6$Sn$_5$ 
\cite{Fisk}. 
This compound shows a very interesting unusual magnetic characteristics.
Especially magnetic and specific heat measurements indicate that 
this compound is very close to ferromagnetic instability and shows 
ferromagnetic ordering after rare earth element doping.
This compound joins other prototype itinerant ferromagnets
ZrZn$_{2}$, Sc$_{3}$In and TiBe$_{2-x}$Cu$_{x}$.
Resistivity and magnetic susceptibility measurements of this compound 
show the Pauli paramagnet displaying Fermi liquid behavior, an 
indication that the Ti in the 4$^{+}$ and not the magnetic 
3$^{+}$ state. The value of $\chi_{0}$ from the 
suscepibility measurements is $1.1\times 10^{-6}$ emu/g,
which is typical value of intermetallic compounds.  
An interesting result comes from the heat capacity measurement.
The heat capacity was measured in the temperature range 
$350mK \leq T \leq 25K$ and straight line which fits to the low 
temperature part of the $\frac{C}{T}$ versus $T^{2}$ gives 
a value of $\gamma(exp)=40$ mJ/K$^{2}$mol. 
Using the values of  $\chi_{0}= 1.1\times 10^{-6}$ emu/g and 
$\gamma$, Drymiotis and coworkers estimated 
the Wilson ratio 
$R_{w}=1.76$ which indicates that  $\beta$-Ti$_6$Sn$_5$ is a
correlated system\cite{Fisk}.
Another interesting results are the effects of doping with rare 
earth elements. 
The small amounts of Ce, La or Co impurities drove
the system to the ferromagnetic (FM) state, with the ordering temperatures
of the order of 150 K. It was shown that the magnetic behavior is intrinsic
to the $\beta$-Ti$_6$Sn$_5$, i.e., it is not the case of magnetic impurities
in non-magnetic host.
For example, the susceptibility measurements on La$_{x}$Ti$_6$Sn$_5$
show ferromagnetic ordering with a critical temperature equal to 
T$_{c}\sim 160 K$.
The constituent elements are non-magnetic, with the exception 
of Ti which in the Ti$^{+}$ configuration carries an effective 
moment of $1.8 \mu_{B}$. 
Susceptibility measurements of the undoped 
$\beta$-Ti$_6$Sn$_5$ show the Pauli paramagnetism, 
which is an indication that the Ti
is in a non-magnetic ground state. 

In this paper we
investigate the electronic and magnetic properties of
$\beta$-Ti$_6$Sn$_5$
based on density functional theory.
In particular we have calculated the bandstructures,
Fermi surfaces and linear specific heat coefficient. 
Large enhancement of the static susceptibility
over its non-interacting value is found due to a peak 
in the density of states at the Fermi level.


\section{Crystal Structure}
The crystal structure of $\beta$-Ti$_5$Sn$_6$  
belongs to the non-symmorphic hexagonal space group P6$_3$/mmc.
This crystal structure is described in detail by Kleinke\cite{kleinke}. 
There are three distinct Sn sites, ((0,0,0), (1/3,2/3, 1/4),(0.795, 
0.590, 1/4)) with site symmeties $\bar{6}m2$, $\bar{6}m2$ and $mm2$ 
respectively and two Ti sites ((1/2, 0,0),(0.165,0.330,1/4)) with 
site symmeties $2/m$ and $mm2$ respectively.
It does not seem to be a simple manner in which to see 
the structure but it appears that the Ti sites form 
groups of tetrahedral along the (001) direction.
$\beta$-Ti$_5$Sn$_6$ is the only known compound which forms 
this structure. The lattice constants a and c are 9.24 $\AA$ and 5.71 $\AA$ 
respectively.
There are two formular units per primitive cell.

Alternatively the crystal structure of $\beta$-Ti$_5$Sn$_6$
can be viewed as ABAB'A stacking of Ti and Ti-Sn layers with 
a separation of $c/4$. The A-plane contains 3 Ti atoms per unit cell
forming a 2D Kagome lattice. The B-plane contains 3 Ti and 5 Sn atoms. 
It is interesting to note that Ti atoms in the B-plane form slightly 
distorted Kagome laticce (the deviations from the ideal bondlenght being
$\pm1 \%$). The B'-layer is the inversion image of the B-layer. The in-plane
nearest-neighbor Ti-Ti distance are 4.61 \AA (A-plane) and 4.61$\pm$0.05 \AA
(B-plane).  
  
\section{Method of Calculations}

We have calculated the electronic structure using the full-potential 
nonorthogonal local-orbital minimum-basis (FPLO) method \cite{Koepernik}  
to solve the Kohn-Sham equations of density functional theory.
Scalar relativistic approximation
and the exchange-correlation functional of Perdew and Wang \cite{perdew}
were use throughout this work.
Ti $3s,3p,4s,4p,3d$ states and Sn $4s,4p,5s,5p,4d$ were included as 
valence states. All lower states were treated as core states.
We included the relatively extended semicore 3s, 3p states of Ti and 
4s, 4p, states of Sn as band states 
because of the considerable overlap of these 
states on nearest neighbors.
This overlap would be otherwise neglected in our FPLO scheme. 
Ti 4p states were added to increase the quality of the basis set. 
The spatial extension of the 
basis orbitals, controlled by a confining potential $(r/r_{0})^4$, was 
optimized to minimize the total energy. The self-consistent potentials were 
carried out on a k mesh of 12 k points in each direction of the Brillouin zone,
which 
corresponds to 133 k points in the irreducible zone.
A careful sampling of the Brillouin zone is necessitated by the 
fine structures in the density of states near Fermi level E$_{F}$.

\begin{figure}
\vskip 5mm
\includegraphics[height=8.5cm,width=8.5cm,angle=-90]{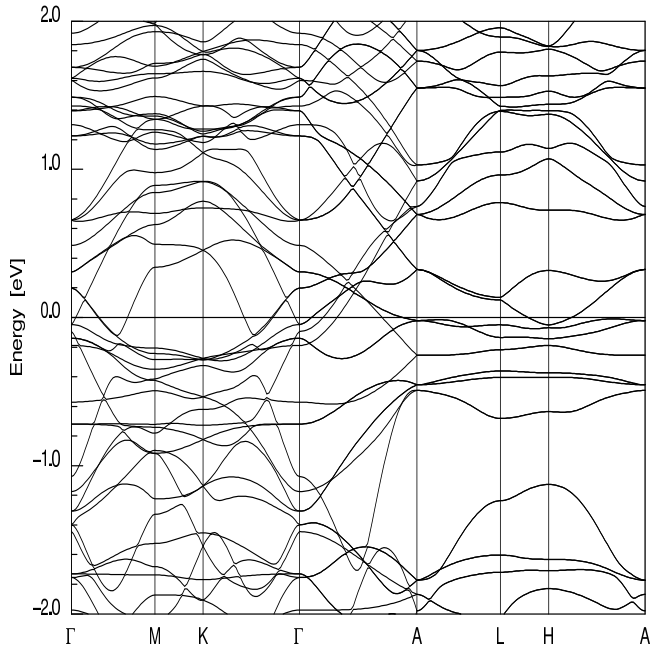}
\caption{The LDA bandstructure of non-magnetic $\beta$-Ti$_{}$Sn$_{5}$
along symmetry lines 
showing that there are several bands with weak dispersion 
being of primarily Ti 3d character near the Fermi level.
} 
\label{band}
\end{figure}

\section{Results and Discussion}

\begin{figure}
\vskip 15mm
\includegraphics[height=8.5cm,width=8.5cm,angle=-90]{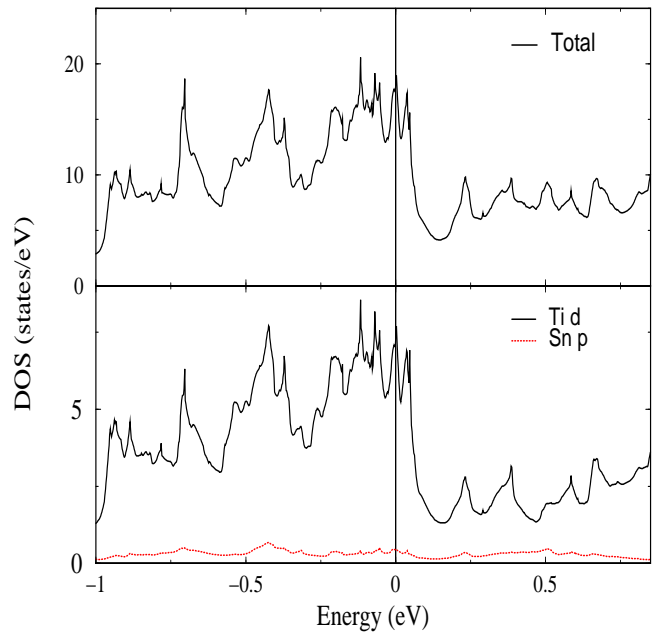}
\caption{Projected density of states of Ti$_{6}$Sn$_{5}$.
Top panel: total density of states.
Bottom panel: projection of the Ti 3d and Sn 5p, showing that Ti 3d 
character dominates the states near the Fermi level. } 
\label{dos}
\end{figure}

In Fig. \ref{band} and Fig. \ref{dos} we show the bandstructure and
the density of state respectively. The bands in the vicinity of the Fermi level
have dominant Ti $3d$ character with the role of Sn to be merely to provide
electrons.
The Sn $5s$ bands lie between -10 eV and -5 eV. 
Around -5 eV there are Ti 4s states, 
above them the bands of Sn $5p$ spread over to the Fermi level.
The Ti $3d$ bands lie between -2.5 eV and 5 eV and they dominate the Fermi 
level.

The bandstructure provides explanation of the rather high
peaks in the density of state, in particular close to the Fermi level.
While the bands in $\Gamma$MK plane exhibit sizable dispersion 
their counterparts
in the ALH plane are rather flat. We offer the following explanation
of this interesting feature. The Ti sublattice consists of Kagome lattices,
well known for a disperssionless band in their spectrum,
coupled along the $c$-axis. Since the Ti-Ti inter-layer (A-A) bond  
is shorter than the Ti-Ti intra-layer distance, the inter-layer coupling
is strong (see also $\Gamma$-A dispersion) and the bandstructure does not
exhibit 2D features. However, the nearest neighbor inter-layer hopping
does not contribute to the dispersion in the ALH plane. This can be shown by
taking the $k_z=\pi/c$ Bloch sums of the relevant $d$ orbitals and considering
the even parity of the $d$ functions. Thus with the contribution of the
main hopping being zero the ALH dispersion reflects the intra-layer hopping.
The corresponding Fermi surfaces (FS) of $\beta$-Ti$_{6}$Sn$_{5}$ 
is shown in Fig. \ref{FS}.
The $\Gamma$ point is located at the center of 
the hexagonal prisim. The large FS in the middle panel of Fig. 
\ref{FS} is due to the Ti 3d states; the contribution of Sn states 
to these sheets is very small.
We also calucated the LSDA magnetic band structures for the 
$\beta$-Ti$_{6}$Sn$_{5}$, in which the exchange spliting is almost negligible. 
Drymiotis {\it et al.}\cite{Fisk} measured the linear specific heat
coefficient for $\beta$-Ti$_6$Sn$_5$ 
of $\gamma$=40 mJ/K$^{2}$ mole(formular unit). 
The calculated value of $N(E_{F})$=41.5 states/Ry/Ti for 
$\beta$-Ti$_6$Sn$_5$
corresponds to
a bare value $\gamma_b$=42.7 mJ/K$^{2}$ mole(formular unit), which 
is in good agreement with the experimental one.  

The presence of an electronic instability is signaled by a divergence of the 
corresponding susceptibility. In the following we study the uniform 
magnetic susceptibility using the method of Janak \cite{jan77}.
The uniform magnetic susceptibility of a metal can be written as

\begin{equation}
\chi=\frac{\chi_{0}}{1-N(E_F)I},
\end{equation}

where the numerator stands for the Pauli susceptibility of a gas of non-interacting
electrons proportional to the density of states at the Fermi level $N(E_F)$,
and the denominator represents the enhancement due to electron-electron
interaction. Within the Kohn-Sham formalism of density functional
theory the Stoner parameter $I$ is related to the second derivative of the
exchange-correlation fuctional with respect to the magnetization density.
We have evaluated, within the density functional theory formalism, the Stoner 
enhancement of the susceptibility 
$\chi=\frac{\chi_{0}}{1-IN(E_{F})}\equiv S\chi_{0} $,
where $\chi_{0}=2\mu_{B}^{2}N(E_{F})$ 
is the non-interacting susceptibility 
and $S$ gives the electron-electron enhancement in terms of the Stoner 
constant $I$. We have calculated $I$ using both the Janak-Vosko-Perdew theory \cite{jan77}
and fixed spin moment calculations\cite{mohn}.
The calculated density of states and Stoner parameter normalized per 
Ti atom are $N(E_F)$=41.5 states/Ry/Ti and $I$=22.9 mRy.
By comparing the calculated value of the density of states with 
the measured susceptibility, a Stoner enhancement S=$[1-IN(E_{F})]^{-1}
\approx 20$ was obtained, indicating $\beta$-Ti$_{6}$Sn$_{5}$ a strongly 
exchange enhanced metal.
The presence of a peak close below the Fermi level
suggest that a very small hole or electron doping can drive system into 
ferromagnetic 
regime. 


The experimental reports \cite{Fisk} show 
that  $\beta$-Ti$_6$Sn$_5$ is located at 
the boundary between non-magnetic  and ferromagnetic ground state. 
This is supported by the large value of $\gamma$, obtained from heat 
capacity measurements which corresponds to the peak at the 
density of states at the Fermi level. The transition to the 
ferromagnetic ground state occurs after a small amount of doping 
on the Ti site. 
Our fixed moment
calculations clearly shows that the $\beta$-Ti$_{6}$Sn$_{5}$
is located at the magnetic instability, which agrees very well with the 
experimental results.
Itinerant magnetism is a 
consequence of an enhanced density of states at the Fermi level, 
consistent with the large values of $\gamma$ in the low temperature 
specific heat data and the magnetic moment is carried 
by the carriers that take part in the transport. 
The enhanced density of states and increase in electronic 
correlations produce the formation of a local moment whose magnitude 
is related to the band structure at the Fermi level. 

\section{Summary}

Our fixed moment
calculations clearly shows that the $\beta$-Ti$_{6}$Sn$_{5}$
is located at the magnetic instability, which agrees very well with the
experimental results.
The calculation indicates that $\beta$-Ti$_{6}$Sn$_{5}$ is very
close to ferromagnetic instability and shows ferromagnetic ordering
after rare earth element doping. Large enhancement of the static
susceptibility over its non-interacting value is found due to
a peak in the density of states at the Fermi level.


\end{document}